\documentclass[aps,prb,twocolumn,showpacs,showkeys,preprintnumbers,groupedaddress,amsmath,amssymb,]{revtex4-1}

\usepackage{graphicx}
\usepackage{dcolumn}
\usepackage{bm}
\usepackage{hyperref}
\usepackage[dvips]{color}

\begin{document}




\title{Crossover of electron-electron interaction effect in Sn-doped indium oxide films}


\author{Yu-Jie Zhang}
\author{Kuang-Hong Gao}
\author{Zhi-Qing Li}
\email[Author to whom correspondence should be addressed. Electronic address: ]{zhiqingli@tju.edu.cn}


\affiliation{Tianjin Key Laboratory of Low Dimensional Materials Physics and
Preparing Technology, Department of Physics, Tianjin University, Tianjin 300072,
China}


\date{\today}

\begin{abstract}
We systematically study the structures and electrical transport properties of a series of Sn-doped indium oxide (ITO) films with thickness $t$ ranging from $\sim$5 to $\sim$53\,nm. Scanning electron microscopy and x-ray diffraction results indicate that the $t\lesssim 16.8$\,nm films are polycrystalline, while those $t\gtrsim 26.7$\,nm films are epitaxially grown along [100] direction. For the epitaxial films, the Altshuler and Aronov electron-electron interaction (EEI) effect governs the temperature behaviors of the sheet conductance $\sigma_\square$ at low temperatures, and the ratios of relative change of Hall coefficient $\Delta R_H/R_H$ to relative change of sheet resistance $\Delta R_\square/R_\square$ are $\approx$2, which is quantitatively consistent with Altshuler and Aronov EEI theory and seldom observed in other systems. For those polycrystalline films, both the sheet conductance and Hall coefficient vary linearly with logarithm of temperature below several tens Kelvin, which can be well described by the current EEI theories in granular metals. We extract the intergranular tunneling conductance of each film by comparing the $\sigma_\square(T)$ data with the predication of EEI theories in granular metals. It is found that when the tunneling conductance is less than the conductance of a single indium tin oxide (ITO) grain, the ITO film reveals granular metal characteristics in transport properties, conversely, the film shows transport properties of homogeneous disordered conductors. Our results indicate that electrical transport measurement can not only reveal the underlying charge transport properties of the film but also be a powerful tool to detect the subtle homogeneity of the film.
\end{abstract}


\maketitle

Tin-doped indium oxide (ITO) is a very interesting and technologically
important transparent conducting oxide (TCO), which simultaneously possesses high electrical conductivity and high optical transparency in the visible light range. It is widely used as transparent electrodes in optoelectronic devices, such as flat panel displays, organic light emitting diodes, solar cells and energy-efficient windows.\cite{Zhang1,zhang2,zhang3,zhang4}  With the rapid development of touch technology, the ITO films are needed to be as thin as possible to improve the performance of small-size touch panels, such as in smart phones and tablet computers.\cite{zhang5,zhang6} However, the change in film thickness may also change the transport property of charge carriers which could in turn affect the sensitivity and stability of the touch devices. For example, it is recently found that when the TCO films are thin enough, their electrical transport properties will be similar to that of granular metals.\cite{zhang7,zhang8} Thus it is necessary to systematically investigate the underlying charge transport mechanisms variation with the thickness of the ultrathin TCO films.

In this Letter, we systematically investigated the temperature behaviors of the electrical conductivity $\sigma$ and Hall coefficient $R_H$ of a series of ITO films with thickness $t$ ranging from $\sim$5 to $\sim$53\,nm. We found that the intergrain electron-electron interaction (EEI) and virtual electron diffusion effects govern the temperature behaviors of $\sigma$ and $R_H$, respectively, for $t<26.7$\,nm film, while the Altshuler-Aronov EEI effect dominates the temperature behaviors of $\sigma$ and $R_H$ of the $t \gtrsim 26.7$\,nm films. Our results indicate that electrical transport measurement can not only reveal the underlying charge transport properties of the films but also be a powerful tool to detect the subtle change from inhomogeneous to homogeneous in film growing process.

Our ITO films were deposited on (100) yttrium stabilized ZrO$_2$ (YSZ) single cryatal substrates by standard rf-sputtering method. The sputtering source was a ceramic Sn-doped In$_{2}$O$_{3}$ target with purity of 99.99\% (the atomic ratio of Sn to In is 1:9). The base pressure of the vacuum chamber was below $8.5\times10^{-5}$\,Pa, and the deposition was carried out in an argon (99.999\%) atmosphere of 0.6\,Pa. During the depositing process, the sputtering power was held at 100\,W and the substrate temperature was fixed at 923\,K. We prepared nine ITO films with different thicknesses by controlling the sputtering time. The thicknesses of the films, ranging from $\sim$5 to $\sim$53\,nm, were determined by the low-angle x-ray diffraction.\cite{zhang7} The surface morphologies of the films were characterized by the scanning electron microscopy (SEM). The crystal structure was determined by the x-ray diffraction (XRD), including normal $\theta$-2$\theta$, $\phi$ and $\omega$ scans.
The electrical conductivity and Hall effect were measured in a physical property measurement system (PPMS-6000, Quantum Design) by the standard four-probe method. In the conductivity measurement process, a magnetic field  of 7\,T and perpendicular to the film plane, is applied to suppress the weak-localization effect.\cite{zhang9,zhang10,zhang11,zhang12,zhang13}
\begin{figure}
\begin{center}
\includegraphics[scale=1.0]{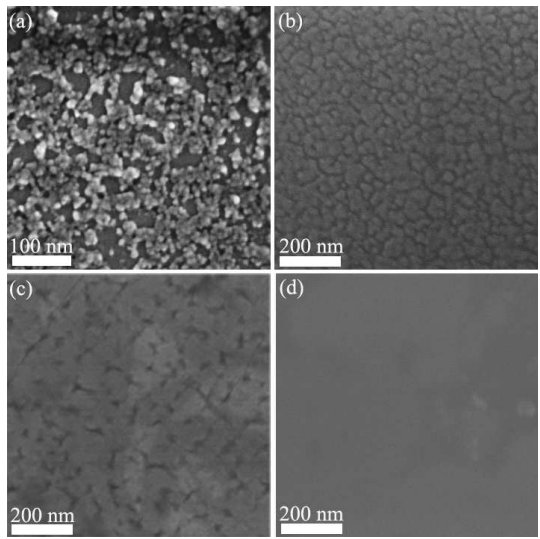}
\caption{SEM micrographs of ITO films with thicknesses of (a) 8.5\,nm, (b) 12.3\,nm, (c) 26.7\,nm, and (d) 35.6\,nm.}\label{FIGSEM}
\end{center}
\end{figure}

Figure~\ref{FIGSEM} presents the SEM images for four representative ITO films with $t$=8.5, 12.3, 26.7, and 35.6\,nm, respectively. For $t\lesssim 8.5$\,nm, the ITO grains have not completely covered the YSZ substrate and the film shows granular-like characteristics in morphology. As for the 12.3\,nm thick film, the substrate is completely covered by the film and the grain boundaries are evident, i.e., the film reveals polycrystalline features. (The microstructure of the 16.8\,nm thick film is similar to that of the 12.3\,nm thick film.) For the 26.7\,nm thick film, the grain boundaries disappear and only some dents or scars, which may be the remnants of the grain boundaries, are left on the surface of the film. When the thickness is further increased to above 35.6\,nm, the dents and scars disappear and the film become uniform in the whole image, i.e., epitaxial growth of ITO film on (100) YSZ may be achieved.
\begin{figure}
\begin{center}
\includegraphics[scale=1]{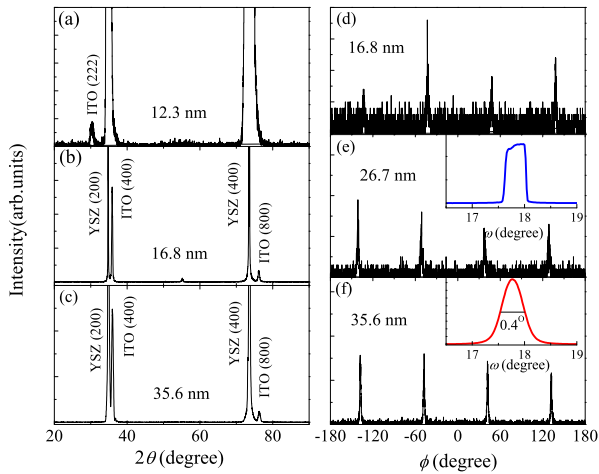}
\caption{(Color online) (a) to (c) Typical $\theta$-2$\theta$ x-ray diffraction patterns for three representative samples. (d) to (f) $\phi$ scan spectra of (211) plane for films with $t=16.8$, 26.7, and 35.6\,nm.  Insets: The $\omega$ scan profiles of the corresponding films shown in the main figures.} \label{FIGXRD}
\end{center}
\end{figure}

Figures~\ref{FIGXRD}(a) to \ref{FIGXRD}(c) present the XRD $\theta$-2$\theta$ scan profiles for three representative ITO films, as indicated. For the $t\lesssim$12.3\,nm films, besides the (400) and (800) diffraction peaks, (222) peak also appear, indicating the films are polycrystalline. This is consistent with the result of SEM. For films with $t\gtrsim$16.8\,nm, only the peaks related to (400) and (800) diffractions appear, indicating the preferred growth orientation is [100] direction [Fig.~\ref{FIGXRD}(b) and \ref{FIGXRD}(c)]. Figures~\ref{FIGXRD}(d) to \ref{FIGXRD}(f) show the $\phi$ scan spectra of (211) plane for three [100]-orientated films. For the $t\gtrsim 35.6$\,nm thick films, four uniform-distribution peaks present in each of the $\phi$ scan patterns, suggesting the films may epitaxially grow on the substrates; the rocking curves ($\omega$ scan spectra) are symmetric and the full widths at half maximum  of the diffraction peaks are $\sim$0.4$^\circ$ [inset of Fig.~\ref{FIGXRD}(f)], which confirms the epitaxial growth. As for the 26.7\,nm thick film, although the $\phi$ scan also possesses fourfold symmetry, the rocking curve is asymmetrical. Hence the film is nearly epitaxial grown and may have many defects. The $\phi$ scan spectrum of the 16.8\,nm thick film loses the fourfold symmetry. Combining to the SEM image, one can obtain that the film  is [100] direction textured polycrystalline film.

\begin{table}
\caption{\label{TableLi} Relevant parameters for the ITO films. $t$ is the mean-film thickness, $R_{\square}(2\,{\rm K})$ is the sheet resistance at 2\,K,  $n^{\ast}$ is the carrier concentration at $T^{\ast}$,  and $T^{\ast}$ is the maximum temperature for $\Delta\sigma\propto\ln T$ law hold. $\sigma_{0}$ and $g_{T}$ are the adjustable parameters in Eq.~(\ref{Eq.(conductance-graular)}), and $c_d$ is the adjustable parameter in Eq.~(\ref{Eq.(HALL)}). In 2D granular arrays, $\sigma_{0}$ represents the sheet conductance without the EEI effec.}
\begin{ruledtabular}
\begin{center}
\begin{tabular}{ccccccc}
  $t$    & $R_{\square}$(2\,K) & $n^{\ast}$  &$T^{\ast}$  &$\sigma_{0}$   & $g_{T}$  &  $c_d$ \\
  (nm)   &($\Omega$) &($10^{27}$ m$^{-3}$) & (K) & ($10^{-3}$ S)  &  &  \\\hline
5.1     &1144   &0.29       &75   &0.97           &6       &1.34\\
7.1     &475    &0.62       &85   &2.22           &13      &1.58\\
8.5     &203    &0.79       &80   &5.06           &29      &1.64\\
12.3    &131    &0.83       &70   &7.75           &46      &1.68\\
16.8    &84     &0.89       &30   &12.2           &67      &1.71\\
26.7    &57     &0.94       &11   &17.8           &112     &-\\
35.6    &43     &0.97       &17   &14.3           &174      &-\\
45.9    &35     &1.10       &17   &28.7           &237      &-\\
52.8    &34     &1.17       &18   &29.5           &261      &-\\
\end{tabular}
\end{center}
\end{ruledtabular}
\end{table}

Recently, it is theoretically found that the influence of EEI effect on the electrical transport properties in granular metals is different to that in homogeneous disordered conductors. Specifically, in granular metals and the strong intergrain coupling limit [$g_{T}\gg$1, where $g_{T}=G_{T}/(2e^{2}/\hbar)$ is the dimensionless intergranular tunneling conductance, $G_{T}$ is the intergrain tunneling conductance, $e$ is the electronic charge, and $\hbar$ is the Planck constant divided by $2\pi$], the intergrain EEI effect governs the temperature behavior of electrical conductivity and leads to a $\ln T$ behavior of $\sigma(T)$, which does not depend on the dimensionality of the samples. Precisely, in the temperature range $g_{T}\delta/k_{B}<T\ll E_{c}/k_{B}$ ($\delta$ is the mean-energy level spacing in the metallic grain,  $k_{B}$ is the Boltzmann constant, and  $E_{c}$ is the charging energy), the electrical conductivity can be written as\cite{zhang14,zhang15,zhang16,zhang17}
\begin{equation}\label{Eq.(conductance-graular)}
\sigma(T)=\sigma_{0}\left[1-\frac{1}{2\pi g_{T}d}\ln\left(\frac{g_{T}E_{c}}{k_{B}T}\right)\right],
\end{equation}
where $\sigma_{0}$ is the conductivity without the EEI effect and $d$ is the dimensionality of the granular array. Whereas in homogeneous disordered metals, the correction to conductivity due to the conventional EEI effect is related to the dimensionality of the system. In two-dimensional (2D) systems, the variation of the sheet conductance is given by\cite{zhang9,zhang18,zhang19}
\begin{equation}\label{Eq.(conductance-homogeneous)}
\Delta \sigma_{\square}(T)=\frac{e^2}{2\pi^{2}\hbar}\left(1-\frac{3}{4}\widetilde{F}\right)\ln\left(\frac{T}{T_0}\right),
\end{equation}
where $T_{0}$ is an arbitrary reference temperature, $\widetilde{F}$ is the electron screening factor.
\begin{figure}[htp]
\begin{center}
\includegraphics[scale=1]{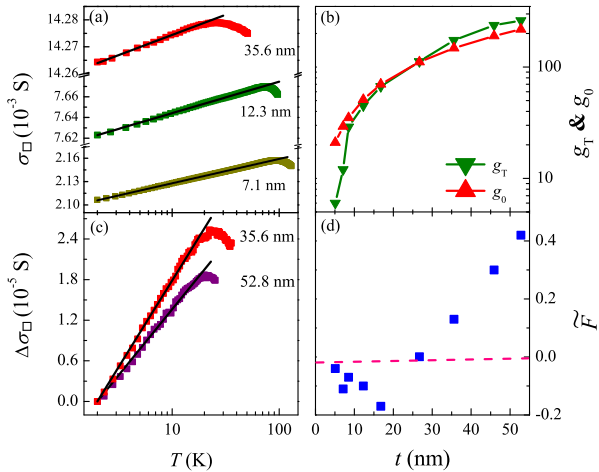}
\caption{(Color online) (a) Sheet conductance  versus temperature for three representative ITO films. The solid straight lines are least-squares fits to Eq.~(\ref{Eq.(conductance-graular)}). (b) Dimensionless intergranular tunneling conductance $g_{T}$ and conductance of a single grain $g_{0}$ as functions of the thickness of the film. The solid curves are only guides to the eye. (c) Change of sheet conductance ${\Delta \sigma}_\square=\sigma_\square (T)-\sigma_\square(2\text{ K})$ as a function of temperature for two homogeneous ITO films. The solid straight lines are least-squares fits to Eq.~(\ref{Eq.(conductance-homogeneous)}). (d) Electron screening factor, obtained by fitting $\Delta\sigma_\square(T)$ data to Eq.~(\ref{Eq.(conductance-homogeneous)}), versus the thickness of the film.} \label{Li-Fig3}
\end{center}
\end{figure}

Figure~\ref{Li-Fig3}(a) shows the variation in sheet conductance $\sigma_{\square}$ with the logarithm of temperature for three representative ITO films, as indicated. Clearly, the conductance varies linearly with $\ln T$ between 2\,K and $T^{\ast}$, where $T^{\ast}$ is the temperature below which the logarithmic law holds. Our experimental $T^{\ast}$ values vary from $\sim$20 to $\sim$85\,K, depending on the samples (see Table I). The linear dependence of $\ln T$ part of $\sigma(T)$ was least-squares fitted to Eq.~(\ref{Eq.(conductance-graular)}), and the fitted results were plotted as straight solid lines in Fig.~\ref{Li-Fig3}(a). In the fitting process, $\sigma_{0}$ and $g_{T}$ are the adjustable parameters and the charging energy is taken to be $E_c \approx 10 k_B T^{\ast}$.\cite{zhang7,zhang17} In addition, considering the film can only be covered by one single layer of ITO grains, we take the dimensionality to be $d=2$. Figure~\ref{Li-Fig3}(b) shows the variation of $g_{T}$ with $t$. Inspection of this figure indicates that $g_{T}$ increases from $\sim$6 to $\sim$260 when $t$ increases from $\sim$5 to $\sim$53\,nm. Theoretically, Eq.~(\ref{Eq.(conductance-graular)}) is derived in the strong coupling limit 1$\ll g_{T}\ll g_{0}$, where $g_{0}=G_{0}/(2e^{2}/\hbar)$ and $G_{0}$ is the conductance of a single metal grain. The condition 1$\ll g_{T}\ll g_{0}$ means that the main contribution to the longitudinal resistance comes from the tunneling barriers between the grains rather than from scattering on impurities inside the grains. When $g_{T}$ is close to $g_{0}$, the system can be viewed as a homogeneous disordered conductor. The resistivity of our 800\,nm thick ITO film grown on (100) YSZ substrate can be as low as $\sim$$5.0\times 10^{-5} \,\Omega$\,cm. Taking this value as the longitudinal resistivity of a single ITO grain and assuming the ITO grains are square in top view, we obtain the $g_{0}$ value for each film. The thickness $t$ dependence of $g_0$ is also shown in Fig.~\ref{Li-Fig3}(b). The difference between $g_T$ and $g_0$ decreases with increasing $t$ but $g_T$ is much less than $g_0$ for those $t\lesssim 7$\,nm films. For those $8.5\lesssim t \lesssim 16.8$\,nm films, $g_T$  gradually approaches $g_0$ with increasing $t$. When $t$ is greater than 26.7\,nm, $g_{T}$ would be greater than $g_{0}$, indicating these films with $t\gtrsim$26.7\,nm have already become homogeneous and Eq.~(\ref{Eq.(conductance-graular)}) fails to describe the $\ln T$ behavior of these films. Thus the homogeneity of the film can be determined by comparing $g_T$ with $g_0$, and the result is consistent with that obtained by SEM and XRD measurements.

We analyze the temperature behavior of these $t\gtrsim 26.7$\,nm films using Altshuler and Aronov EEI theory.\cite{zhang9,zhang18} For the thickest film ($t=52.8$\,nm), the electron thermal diffusion length is estimated to be $L_{T}=\sqrt{\hbar D/(k_{B}T)} \approx 91$\,nm at 100 K.\cite{zhang20} Hence our ITO films are 2D with regard to EEI effect. Figure~\ref{Li-Fig3}(c) shows the change of sheet conductance versus the logarithm of temperature for the 35.6 and 52.8\,nm thick films and the solid straight lines are least-squares fits to Eq.~(\ref{Eq.(conductance-homogeneous)}). Here the reference temperature $T_0$ is taken as 2\,K. Clearly, the experimental $\Delta\sigma_\square(T)$ data can be well described by Eq.~(\ref{Eq.(conductance-homogeneous)}),
and the fitted parameter $\widetilde{F}$ increases with increasing $t$ [Fig.~\ref{Li-Fig3}(d)]. Theoretically, the screening effect of electrons increases with increasing electron density.\cite{zhang21} Inspection of Table~\ref{TableLi} indicates that the carrier concentration indeed increases with increasing $t$. Hence the observed $t$ dependence of $\widetilde{F}$ is reasonable and self-consistent.
For those $t<26.7$\,nm films, the theoretical predictions of Eq.~(\ref{Eq.(conductance-homogeneous)}) are also coincident with the $\sigma_\square(T)$ data, however, the obtained $\widetilde{F}$ values are negative [see Fig.~\ref{Li-Fig3}(d)]. Since the conventional EEI theory requires that 0$\lesssim\widetilde{F}\lesssim$1, our results indicate that the $\ln T$ behavior of $\sigma(T)$ for the $t<26.7$\,nm films can not ascribe to the Altshuler and Aronov EEI effect, and are specific to granular system. We thus observed the crossover of the EEI effect from intergrain to Altshuler and Aronov types in ITO films by tuning the thicknesses of the films.

According to the recent theories,\cite{zhang22,zhang23} EEI effect due to the presence of granularity also affects the temperature behavior of the Hall coefficient, which is also different to that of homogeneous disordered conductors. More precisely, when the virtual electron diffusion effect is considered, the change of Hall coefficient can be written as\cite{zhang22,zhang23}
\begin{equation}\label{Eq.(HALL)}
\Delta R_{H}=-\frac{c_{d}}{4\pi n^{\ast}e g_T} \ln \left(\frac{T_0}{T}\right),
\end{equation}
where $n^{*}$ is the effective carrier concentration, $c_{d}$ is a numerical lattice factor. Eq.~(\ref{Eq.(HALL)}) is valid in the temperature range $g_{T}\delta/k_{B}\lesssim T\lesssim\text{min}(g_{T}E_{c},E_{\text{Th}})/k_{B}$, where $E_{\text{Th}}$ is the Thouless energy. The minus on the right hand side of Eq.~(\ref{Eq.(HALL)}) represents the main charge carrier is electron, for $e$ is positive by definition.
\begin{figure}[htp]
\begin{center}
\includegraphics[scale=1]{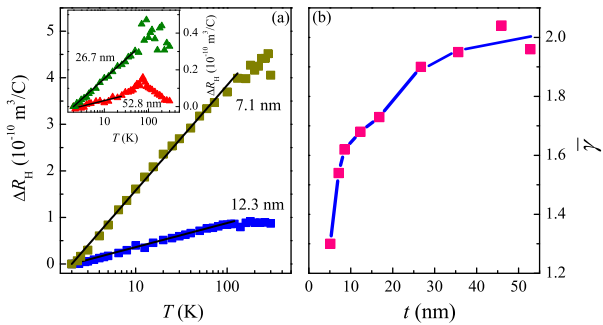}
\caption{(Color online) (a) Change of Hall coefficient $\Delta R_{H}=R_{H}-R_{H}(2\,{\rm K})$ versus temperature for two representative ITO granular films. The solid straight lines are least-squares fits to Eq.~(\ref{Eq.(HALL)}). Inset: Change of Hall coefficient $\Delta R_{H}$ versus temperature for two representative homogeneous ITO films. The solid straight lines  are only guides to the eye. (b) $\bar{\gamma}$, the average value  of the ratio $[\Delta R_{H}/R_{H}(2\,{\rm K})]/[\Delta R_{\square}/R_{\square}(2\,{\rm K})]$, as a function of the thickness of the film. The solid curve is only guide to the eye.}
\label{Li-Fig4}
\end{center}
\end{figure}

Figure~\ref{Li-Fig4}(a) shows the change of Hall coefficient $\Delta R_{H}=R_{H}-R_{H}(2\,{\rm K})$ as a function of logarithm of temperature for two representative granular films, as indicated. The Hall coefficients are negative for all films, indicating the charge carrier is electron.  Inspection Fig.~\ref{Li-Fig4}(a) indicates the Hall coefficients of the films vary linearly with $\ln T$ from 2\,K up to $T_{\text{max}}$, where $T_{\text{max}}$ is the temperature below which the logarithmic law holds. The values of $T_{\text{max}}$ vary from $\sim$85 to $\sim$130\,K for our granular ITO films. The $\Delta R_{H}$ data were least-squares fitted to Eq.~(\ref{Eq.(HALL)}), and the fitted results were plotted as straight solid lines in Fig.~\ref{Li-Fig4}(a). (In the fitting process, the $n^{\ast}$ value was taken the average value of the carrier concentration near $T_{\rm max}$, and $c_d$ is the only adjustable parameter.)  From this figure, one can see that Eq.~(\ref{Eq.(HALL)}) can well describe our experimental $\Delta R_{H}$ in a considerably wide temperature range. The fitted parameter $c_{d}$ is listed in Table I. For the homogeneous ITO films, the changes of Hall coefficients $\Delta R_{H}=R_{H}-R_{H}(2\,{\rm K})$ also vary linearly with $\ln T$ below $\sim$30\,K, as shown in the inset of Fig.~\ref{Li-Fig4}(a).

We extract the ratio ($\gamma$) of relative change of Hall coefficient $[R_{H}(T)-R_{H}(2\,{\rm K})]/R_{H}(2\,{\rm K})$ to that of the sheet resistance $[R_\square(T)-R_\square(2\,{\rm K})]/R_\square(2\,{\rm K})$ for each film. We found that $\gamma$ is almost a constant for each film as $T$ varies from 2\,K to $T^{\ast}$ ($2\,{\rm K}<T<T^{\ast}$). Figure~\ref{Li-Fig4}(b) shows the variation in the $\bar{\gamma}$ with $t$ for the films, where $\bar{\gamma}$ is average value of $\gamma$ between 2\,K and $T^{\ast}$. Inspection of Fig.~\ref{Li-Fig4}(b) indicates that the $\bar{\gamma}$ values vary from 1.90 to 2.04 ( $\bar{\gamma}\approx 2$) for the homogeneous (epitaxial) ITO films, while the values of $\bar{\gamma}$ vary from 1.73 to 1.30 as $t$ decreases from 16.8 to 5.1\,nm for those inhomogeneous (granular) films. According to Altshuler, Aronov and Lee,\cite{zhang9,zhang10,zhang18,zhang19} the conventional EEI effect also causes a correction to the Hall coefficient, which is related to the correction to the resistance. More precisely, the relation between the changes of the Hall coefficient and the sheet resistance is given by
$\Delta R_{H}/R_{H}$=2$\Delta R_\square/R_\square$.\cite{zhang9,zhang10,zhang18,zhang19} Thus the result $\bar{\gamma}\approx2$ for each homogeneous ITO film agrees well with the predication of Altshuler \emph{et al}'s theory. In fact, such kind of consistency in the ratio $\gamma$ between the experimental result and the theoretical predication is seldom observed in other systems although this predication was proposed three decades ago. Ref.~\onlinecite{zhang24} is considered as the first paper to demonstrate the predication. However the authors found that the ratio $\gamma$ tend to be 2 only in the limit $R_\square\rightarrow 0$ and $H\rightarrow 0$, where $R_\square$ is the sheet resistance of the electron gas in Si metal-oxide-semiconductor field effect transistors and $H$ is the magnetic field applied in Hall effect measurement. Successively, it is found that the ratio $\gamma$ is $\sim$1.5 in 6.2\,nm thick Au film,\cite{zhang25} and 1.4 in 7.5 to 14\,nm thick GeSb$_2$Te$_4$ films.\cite{zhang26} Hence ITO thin film is an ideal model system to quantitatively test the quantum electron transport theories, which may be related to the free-electron-like energy bandstructure of this material.\cite{zhang3,zhang27,zhang28,zhang29,zhang30,zhang31}
According to Eq.~(\ref{Eq.(conductance-graular)}) and Eq.~(\ref{Eq.(HALL)}), the relative changes of the Hall coefficient and resistance should obey $\Delta R_{H}/R_{H}\approx(c_{d}d/2)\Delta R_{\square}/R_{\square}$ in the temperature range $g_{T}\delta/k_{B}<T\ll E_{c}/k_{B}$.\cite{noteHall} For the 2D ITO granular arrays, the theoretical value of the ratio would be $\bar{\gamma}\approx c_d$. Checking Fig.~\ref{Li-Fig4}(b) and Table I, one can readily see that the value of $\bar{\gamma}$ is almost identical to the value of $c_{d}$ for those $t \lesssim 16.8$\,nm films. This result in turn provides strong evidence for the validity of Eqs.~(\ref{Eq.(conductance-graular)}) and (\ref{Eq.(HALL)}).

In summary, we deposit a series of ultrathin ITO films on (100) YSZ single crystal substrates by rf-sputtering method. The $t\lesssim 16.8$\,nm films are polycryatalline films while those $t\gtrsim 26.7$\,nm films are epitaxially grown on the substrates. The temperature behaviors of sheet conductance and Hall coefficient of the epitaxial film can be well described by Altshuler and Aronov EEI theory, and the ratios of relative change of Hall coefficient $\Delta R_H/R_H$ to relative change of resistance $\Delta R_\square/R_\square$ are $\approx$2. For the  $t\lesssim 16.8$\,nm films, the temperature behaviors of sheet conductances and Hall coefficients can only be quantitatively described by the current theory of EEI effect in the presence of granularity. Comparing the intergranular tunneling conductance deduced from the current EEI theory with the conductance of a single ITO grain, one can precisely determine the homogeneity of the ITO film. Thus the electrical transport measurement is also a powerful tool to detect the subtle change of the homogeneity of the film.

The authors are grateful to Professor J. J. Lin for valuable discussion. This work was supported by the National Natural Science Foundation of China (NSFC) through Grant No. 11174216, Research Fund for the Doctoral Program of Higher Education through Grant No. 20120032110065 (Z.Q.Li.).


\begin{thebibliography}{00}\label{sec:TeXbooks}
\bibitem{Zhang1}J. Ederth, P. Johnsson, G. A. Niklasson, A. Hoel, A. Hult{\aa}ker, P. Heszler, C. G. Granqvist, A. R. van Doorn, M. J. Jongerius, and
D. Burgard, Phys. Rev. B 68, 155410 (2003).
\bibitem{zhang2}A. Gondorf, M. Geller, J. Wei{\ss}on, A. Lorke, M. Inhester, A. Prodi-Schwab, and D. Adam, Phys. Rev. B 83, 212201 (2011).
\bibitem{zhang3}P. D. C. King and T. D. Veal,  J. Phys.: Condens. Matter 23, 334214 (2011).
\bibitem{zhang4}J. J. Lin and Z. Q. Li, J. Phys.: Condens. Matter 26, 343201 (2014).
\bibitem{zhang5}H. Tian, D. Xie, Y. Yang, T. L. Ren, Y. F. Wang, C. J. Zhou, P. G. Peng, L. G. Wang, and L. T. Liu, Appl. Phys. Lett. 99, 043503 (2011).
\bibitem{zhang6}J. H. Lu, B. Y. Chen, and C. H. Wang, J. Vac. Sci. Technol. A 32, 02B107 (2014).
\bibitem{zhang7}Y. J. Zhang, Z. Q. Li, and J. J. Lin, Phys. Rev. B 84, 052202 (2011).
\bibitem{zhang8}Y. Yang, Y. J. Zhang, X. D. Liu, and Z. Q. Li, Appl. Phys. Lett. 100, 262101 (2012).
\bibitem{zhang9}B. L. Altshuler, D. Khmelnitzkii, A. I. Larkin, and P. A. Lee, Phys. Rev. B 22, 5142 (1980).
\bibitem{zhang10}P. A. Lee and T. V. Ramakrishnan, Rev. Mod. Phys. 57, 287 (1985).
\bibitem{zhang11}G. Bergmann, Phys. Rep. 107, 1 (1984).
\bibitem{zhang12}G. Bergmann, Int. J. Mod. Phys. B 24, 2015 (2010).
\bibitem{zhang13}J. J. Lin and J. P. Bird, J. Phys.: Condens. Matter 14, R501 (2002).
\bibitem{zhang14}K. B. Efetov and A. Tschersich, Phys. Rev. B 67, 174205 (2003).
\bibitem{zhang15}K. B. Efetov and A. Tschersich, Europhys. Lett. 59, 114 (2002).
\bibitem{zhang16}I. S. Beloborodov, K. B. Efetov, A. V. Lopatin, and V. M. Vinokur, Phys. Rev. Lett. 91, 246801 (2003).
\bibitem{zhang17}Y. C. Sun, S. S. Yeh, and J. J. Lin, Phys. Rev. B 82, 054203 (2010).
\bibitem{zhang18}B. L. Altshuler, A. G. Aronov, and P. A. Lee, Phys. Rev. Lett. 44, 1288 (1980).
\bibitem{zhang19}B. L. Altshuler and A. G. Aronov, in\emph{ Electron-Electron Interactions in Disordered Systems}, edited by A. L. Efros and M. Pollak (Elsevier, Amsterdam, 1985).
\bibitem{zhang20}In this work, the electron diffusion constant ($D$) value  was calculated by using an effective electron mass $m^{*}\thickapprox 0.4m_{e}$ (Ref.~\onlinecite{Zhang1}), where $m_{e}$ is the free electron mass.
\bibitem{zhang21}C. Kittel, \textit{Introduction to Solid State Physics}, 7th ed, (John Wiley and Sons, Inc., 1996), p. 280.
\bibitem{zhang22}M. Yu. Kharitonov and K. B. Efetov, Phys. Rev. Lett. 99, 056803 (2007).
\bibitem{zhang23}M. Yu. Kharitonov and K. B. Efetov, Phys. Rev. B 77, 045116 (2008).
\bibitem{zhang24}D. J. Bishop, D. C. Tsui, and R. C. Dynes, Phys. Rev. Lett. 46, 360 (1981).
\bibitem{zhang25}G. Bergmann, Solid State Commun. 49, 775 (1984).
\bibitem{zhang26}N. P. Breznay, H. Volker, A. Palevski, R. Mazzarello, A. Kapitulnik, and M. Wuttig, Phys. Rev. B 86, 205302 (2012).
\bibitem{zhang27}I. Hamberg, C. G. Granqvist, K. F. Berggren, B. E. Sernelius, and L. Engstr\"{o}m, Phys. Rev. B 30, 3240 (1984).
\bibitem{zhang28}O. N. Mryasov and A. J. Freeman, Phys. Rev. B 64, 233111 (2001).
\bibitem{zhang29}H. Odaka, Y. Shigesato, T. Murakami, and S. Iwata, Japan. J. Appl. Phys., Part 1 40, 3231 (2001).
\bibitem{zhang30}Z. Q. Li and J. J. Lin, J. Appl. Phys. 96, 5918 (2004).
\bibitem{zhang31}N. Kikuchi, E. Kusano, H. Nanto, A. Kinbara, and H. Hosono, Vacuum 59, 492 (2000).
\bibitem{noteHall} Since the EEI corrections to $R_H$ and $\sigma$ are both perturbation terms, the $R_H$ and $\sigma$ in denominators are taken to be $1/(n^{\ast}e)$ and $\sigma_{0}$, respectively.
\end{thebibliography}
\end{document}